\documentstyle[aps,prl,multicol,psfig]{revtex}

\newcommand{\ad}{a^\dagger} 
\newcommand{\Tr}{\mbox{Tr}} 
\newcommand{\id}{{\sf 1 \hspace{-0.3ex} 
\rule{0.1ex}{1.52ex}\rule[-.02ex]{0.3ex}{0.1ex} }}

\newcommand{\bra}[1]{\langle#1|} 
\newcommand{\ket}[1]{|#1\rangle} 
\newcommand{\braket}[2]{\langle#1|#2 \rangle} 
\newcommand{\ketbra}[2]{|#1\rangle\!\langle #2|} 
\newcommand{\ketbrad}[1]{|#1\rangle\!\langle #1|} 
\newcommand{\eqref}[1]{(\ref{#1})}
\newcommand{\mb}[1]{\mbox{\boldmath$#1$}}
\newcommand{\mc}[1]{{\mathcal #1}}
\newcommand{\ea}{\emph{et al.}}

\begin{document}
\title{Generalized measurements by linear elements}
\author{John Calsamiglia}
\address{Helsinki Institute of Physics, PL 64, 
FIN-00014 Helsingin yliopisto, Finland}
               
\date{Received \today}
\maketitle
\begin{abstract} 
I give a first characterization  of the
class of  generalized measurements that can be exactly realized 
on a pair of qudits encoded in indistinguishable particles,
by  using only linear elements and particle detectors.
Two immediate results follow from this characterization.
(i) The Schmidt number of each POVM element cannot exceed the number of initial 
particles. This rules out any possibility of performing perfect
Bell-measurements for qudits. (ii) The maximum probability of 
performing a generalized incomplete Bell-measurement is $\frac{1}{2}$.
\end{abstract}
\begin{multicols}{2}

\section{Introduction}
In the last years there has appeared very important contributions 
to the field of quantum information processing with linear elements 
(see below).
Linear elements  provide the means to 
exploit symmetry
and interference effects associated with indistinguishable particles.
This raises many interesting questions from the fundamental point of 
view, but it is also highly relevant technologically.
Quantum information processing has a wide range of striking applications
\cite{nielsen}.
Many of these applications have been first implemented in optical 
systems, where the lack of interaction at the single 
photon level \cite{ppint} makes  the indistinguishability 
of the photons a crucial feature. 
Photons are ideal qudit ($d$-dimensional counterpart of a qubit)
carriers,
because of their low decoherence rates,
and linear optical elements are extremely simple devices which 
allow one to perform certain operations on the encoded 
photons in a controlled fashion.
It is therefore very desirable to know
the capabilities and limitations of
those operations.
In a recent work E. Knill \emph{et al.}\cite{knill01} make an important 
breakthrough in this direction
showing that fault-tolerant computation with linear 
 optics is in principle possible. 
Starting from the idea of teleportation of quantum 
gates\cite{gottesman},
they develop a method to perform any quantum operation
with a probability that \emph{asymptotically}
approaches unity with a growing number of highly entangled auxiliary photons.
The preparation of the required auxiliary states is, however, far beyond
the current technological possibilities.
Moreover, their work makes a big step forward by
presenting a proof of principle, but it does not elucidate
the role played by the particle symmetry and indistinguishability,
nor does exclude the idea that for specific applications
one can find simpler protocols\cite{ralph01} 
with less technological restrictions.
 For example, recent research shows
how to purify entangled photons in noisy communication channels\cite{pan01} 
or from parametric down-conversion sources\cite{simon01},
reject bit flip errors in quantum communication\cite{bouw01}, perform 
optimal unambiguous state discrimination\cite{huttner}, and
efficiently eavesdrop a quantum key distribution\cite{calsamig01b}, by 
using only a few beam splitters and particle detectors.
Thus, the prospect of new applications and the need for a deeper
understanding warrants further research on the power of linear 
elements.

In this communication I address the question 
of performing  generalized measurements on indistinguishable particles
by linear elements and particle detectors.
For brevity, I term the
qudits 
that are encoded in indistinguishable particles \emph{i-qudits},
 \emph{b-qudits} for bosons, and \emph{f-qudits} for fermions.
At first sight it seems that linear elements cannot realize 
non-trivial generalized measurements, for they are unable to
provide interaction between the particles.
However, this argument has its roots in the concept of locality,
which becomes vague when dealing with indistinguishable particles.
Assigning a notion of locality to the
i-qudits is only possible if each i-qudit occupies a different set of 
modes, so that Hilbert space of the whole system is the tensor product
of the local Hilbert spaces of the individual qudits.
At a descriptive level there is essentially no difference
between i-qudits and qudits.
Things change dramatically when one explores 
the technological possibilities of
performing quantum operations on the i-qudits. 
As soon as the mixing of modes becomes possible the notion
of locality vanishes, and realizing some non-trivial
measurement becomes possible.

Until now, the measurement on i-qudits
has only been studied in the context of unambiguous discrimination
of given set of states. In Refs.\cite{lutvaid,dusek01} 
the impossibility of performing a complete Bell-measurement on two 
i-qudits was proved and the optimum efficiency
of the incomplete Bell-measurement was given in Ref.\cite{calsamig01}.
Later Carollo \ea\cite{carollo01} showed that for a particular set of states,
which are product states, discrimination without error is also impossible.
The general approach in these papers was to feed the linear device 
with the states to be discriminated and check under what conditions  
the particle detectors at the output produce different ``click'' 
combinations that could identify the input state.
Here I will adopt a different approach.
Given that the measurement outcomes are of a known form,
namely a ``click'' pattern,  we will find the POVM (see below)
on the input i-qudits induced by this type of measurement.
This approach is much more general and sets a suitable 
framework to arrive at the full characterization of the class of
POVMs implemented by linear elements.

\section{i-Qudits}\label{sec:form}
An arbitrary one qudit state 
$\ket{\mb{\alpha}}=\sum_{i=1}^d\alpha_{i}\ket{i}$
is encoded in a single excitation, say, a photon,
occupying $d$ field modes, 
$\ket{\mb{\alpha}}= \sum_{i=1}^d\alpha_{i}\ad_{i}\ket{0}$.
Here $\ket{0}$ denotes the vacuum state and
$\ad_{i}$ are bosonic (fermionic) creation operators.
 Whenever needed, I will give the results corresponding to
each of the particle statistics.
In order to encode a two qudit state $\ket{C}=\sum_{i,j=1}^d 
C_{ij}\ket{i}\ket{j}$ a second particle is used occupying
$d$ extra modes $\{\ad_{d+1},\ldots,\ad_{2d}\}$,
$\ket{C}=\sum_{i,j=1}^d C_{ij}\ad_{i}\ad_{d+j}\ket{0}$.
Any two-boson (fermion) state can be defined with 
 a bilinear form,
\begin{equation}
    \ket{\Psi} = \sum_{i,j=1}^n N_{i j}\ad_{i}\ad_{j} =
    {\bf a}^T N {\bf a} \ket{{\bf 0}}\mbox{,} 
    \label{eq:min}
\end{equation}
where $N$ is a $ n\!\times\! n$ symmetric (antisymmetric) matrix and
${\bf a}=(a_{1}^\dagger,\ldots,a_{n}^\dagger)^T$. 
In particular, the bilinear form of the 
two i-qudit state $\ket{C}$ is
\begin{equation}
  N=\frac{1}{2}\left( 
\begin{array}{c|c} 
\begin{array}{cc} 0 & C \\ {\scriptscriptstyle ^{(-)}}C^T & 0 \end{array}
   & \begin{array}{ccc}  &  \\  & 0   \\ &   \end{array} \\
\hline
0 & 0
\end{array} 
\right). 
    \label{eq:N}
\end{equation}
The $d\!\times\! d$ matrix $C$ is defined using the correspondence between 
the state vectors $\ket{C}=\sum_{i,j=1}^d  C_{i j}\ket{i}\ket{j}$ and 
the $n\!\times\! n$ complex matrix $C$ with elements $C_{i j}$. 
Matrix analysis theory \cite{horn91} renders
this representation into a very convenient one for studying bipartite quantum 
systems \cite{dariano}. Some useful
relations between both representations are
\begin{eqnarray}
    A\otimes B\ket{C} & = & \ket{ACB^T}\mbox{,}
    \label{eq:abc}  \\
    \braket{A}{B} & = & \Tr(A^\dagger B) \mbox{,}  \label{eq:innerprod}\\
    \Tr_{1}(\ketbra{A}{B}) & = & AB^\dagger \mbox{ and } 
    \Tr_{2}(\ketbra{A}{B}) = A^TB^*\mbox{.}
    \label{eq:red}  
\end{eqnarray}
I write matrices and vectors in the canonical basis. Thereby,
the correspondence between matrices and bipartite 
state vectors (which is obviously basis dependent) 
is always well defined.
 
The action of the linear elements is defined by a linear 
mapping of the input creation operators
$\{a_{1}^\dagger\ldots\ad_{n}\}$ to  the
output creation operators $\{c_{1}^\dagger\ldots c_{n}^\dagger\}$
\begin{equation} 
    c_{i}^\dagger =
    \sum_{j=1}^{n} U^\dagger_{i j} a_{j}^\dagger\mbox{.}
    \label{eq:linComb}
\end{equation}
To implement this operation 
one can use a series of beam splitters and phase 
shifters \cite{reck94a} or multiports\cite{howell00}.

\section{Generalized measurements on i-qudits}\label{sec:2povm}
A generalized measurement is described by a positive
operator valued measure (POVM) \cite{peres93} given by a collection of
positive operators $F_k$ with $\sum_k F_k = \id $. Each operator 
$F_k$ corresponds to one classically distinguishable measurement 
outcome (e.g. a given combination of ``clicks'' in the output 
detectors).
The probability $p(k|\rho)$ for the outcome
$k$ to occur, conditioned to an input density matrix
$\rho$,  is  $p(k|\rho) = \Tr(\rho F_k)$.  

If we send an i-qudit $\ket{\mb{\alpha}}$ through a linear device,
the state in the output is $\ket{\mb{\alpha}_{out}}=\sum_{j=1}^{n}U_{ji} 
        \alpha_{j}c^\dagger_{i}\ket{0}=\ket{U^T\mb{\alpha}}$,
where the vector $\mb{\alpha}$ is padded with 
extra zeros whenever $n>d$. 
Notice that the number of modes involved 
in the transformation can be larger than the number of modes occupied 
by the i-qudit.  This provides a straightforward extension of our input 
Hilbert space, $\mc{H}_{1}\oplus\mc{H}_{1'}$ where $\mc{H}_{1}$ is the 
i-qudit Hilbert space and $\mc{H}_{1'}$
is spanned by single particle states occupying modes 
$\{\ad_{d+1}\ldots\ad_{n}\}$. The particle detectors in the
output modes effectively perform a von Neumann measurement in the canonical base
of the extended Hilbert space. According to Neumark's theorem
\cite{peres93} \emph{any} POVM can be realized following this 
prescription, i.e. unitary transformation of the initial states in an 
extended Hilbert space followed by a projection measurement.
Explicitly, the event of one ``click'' in mode $c_{i}$ is 
associated to the POVM element $F^{i}$,
\begin{eqnarray}
  && \Tr(F^{i}\ketbrad{\mb{\alpha}})= p(i|\mb{\alpha})  =  |\bra{0} 
    c_{i}U\ket{\mb{\alpha}}|^2  \nonumber\\ 
    & &=\Tr(\ket{\mb{v_{i}}}\braket{\mb{v_{i}}}{\mb{\alpha}}
    \bra{\mb{\alpha}}) \mbox{  } \forall\mb{\alpha} \;
    \longrightarrow \;  F^i=\ketbrad{\mb{v_{i}}}\label{eq:1povm}
\end{eqnarray}
where the $d$-dimensional vector $\mb{v_{i}}=(U^{\dagger}_{1i},\ldots, 
U^{\dagger}_{di})^T$. 
Naturally this generalized measurement is destructive in the sense that
the detectors absorb the measured particle.
Nevertheless many protocols in quantum information conclude
with a measurement which can be destructive; thus, by allowing 
postselection, one can use linear elements to perform
any generalized measurement on a single qudit within the protocol.
This is the reason why protocols such as optimal unambiguous state 
discrimination\cite{huttner}, or some particular entanglement transformation 
\cite{zhang01} can be successfully carried out in optical 
implementations. 

As already mentioned, the situation is quite different in the two 
i-qudits case. A two i-qudit state $\ket{C}$ described, according to 
Eq.~\eqref{eq:N}, by a matrix $N$ will be transformed to 
a two-particle state with the following matrix representation in terms 
of the output modes,
\begin{equation}
   \ket{C}={\bf c}^T M {\bf c} \ket{{\bf 0}} \; \mbox{ with }  
   M=U^T N U\mbox{,}
    \label{eq:mout}
\end{equation}
where ${\bf c}=(c_{1}^\dagger,\ldots,c_{n}^\dagger)^T$.
Notice that the only mode transformations 
which leave the i-qudits Hilbert space invariant are
\begin{displaymath}
    U_{\rm{sep}}=\left(
    \begin{array}{ccc}
        U_{1} & & \\
         & U_{2} & \\ 
	& & U_{3}
    \end{array}
    \right)  
    \mbox{,}\;
     U_{\rm{sw}}=
      \left(\begin{array}{cc|c}
             0& \id_{d} & \\
             \id_{d}& 0 & \\ 
	     \hline
           & & \id_{n-2d}
          \end{array}\right) 
	  \nonumber 
\end{displaymath}
and compositions of both ($U_{1}$ and $U_{2}$ are 
$d\!\times\!d$ unitary matrices). From Eq.~\eqref{eq:abc} it follows that
the first transformation corresponds to a separable operation
in the i-qudits Hilbert space $U_{1}\!\otimes\!U_{2}\ket{C}$,
while the second transforms the i-qudit $\ket{C}$ to
${\scriptstyle (-)}\ket{C^T}$,
i.e. performs the non-separable swap operation. 
I now define,
\begin{equation}
    U=\left(\begin{array}{ccc}
           & A &\\
	   \hline
           & B  & \\ 
	     \hline
           & D & 
          \end{array}\right)\mbox{,}
    \label{eq:Uabc}
\end{equation}
where $A$ and $B$ are $d\!\times\!n$ matrices.
From Eqs.~\eqref{eq:N} and ~\eqref{eq:mout} it is clear that the output state 
$M$ will not depend on the values of matrix elements of $D$,
i.e. on how the initially unoccupied modes transform.
Now we are in position to calculate the resulting POVM on the i-qudits
when particle detectors are placed in the output modes.
Linear elements preserve the number of particles;
therefore each measurement outcome is associated to the absorption
of two particles at modes $c_{i}$, $c_{j}$. Given an arbitrary two i-qudit
state $\ket{C}$, the probability amplitude
of such an event is (for $i\neq j)$
\begin{eqnarray}
   && \bra{0}c_{i}c_{j}\ket{C}= 
    \bra{0}M_{kl}c_{i}c_{j}c^{\dagger}_{k}c^{\dagger}_{l}\ket{0}=
    2\bra{i} M \ket{j}
    \nonumber  \\
    & &= \bra{i} A^T C B \pm B^T C^T A \ket{j}=
    \Tr(A^T C B (\ketbra{i}{j}\pm\ketbra{j}{i})) \nonumber\\ 
   & &= \Tr(C B(\ketbra{i}{j}\pm\ketbra{j}{i}) A^T )=\Tr(C 
   P^{ij}{}^\dagger)\mbox{,}   
    \label{eq:deriv}
\end{eqnarray}
where  the $d\!\times\!d$ matrix $P^{ij}$ is defined as
\begin{equation}
P^{ij}  =  A^{*} \Delta^{ij} B^\dagger \;\mbox{with}  
\; \Delta^{ij}_{k l}= 
\delta_{ki}\delta_{lj}\pm\delta_{kj}\delta_{li}\mbox{,}
\label{pij}
\end{equation}
and the $+(-)$ refers to the b-qudits (f-qudits) result.
In the bosonic case a normalizing factor $\frac{1}{\sqrt{2}}$
should be added to Eqs.~\eqref{eq:deriv} and \eqref{pij}
when $i=j$. The isomorphism between bipartite pure states 
and matrices assigns the state 
$\ket{P^{ij}}=\sum_{i,j}^{n}P^{ij}\ket{i}\ket{j}$ to 
this matrix. Moreover, Eq.~\eqref{eq:innerprod} 
allows us to write the probability amplitude as 
\begin{equation}
    \bra{0}c_{i}c_{j}\ket{C}=\braket{P^{ij}}{C}\mbox{.}
    \label{eq:pramp}
\end{equation}
The relation of the probability amplitude of this event 
to the corresponding POVM element $F^{ij}$ results in
\begin{eqnarray}
   && \Tr(\ketbrad{C} F^{ij}) = p(ij|C) = |\braket{P^{ij}}{C}|^2  
    \nonumber\\
    &&= \Tr(\ketbrad{C}\ketbrad{P^{ij}}) 
    \longrightarrow F^{ij}= \ketbrad{P^{ij}}\mbox{.}
    \label{eq:rk}
\end{eqnarray}
Making use of Eq.~\eqref{eq:abc}, we arrive at the central result,
\begin{eqnarray}
    \ket{P^{ij}} & = & \sqrt{2} A^*\!\otimes\!B^* \ket{\psi^{ij}} \;\mbox{or}
    \label{eq:pije}  \\
    \ket{P^{ij}} & = & \ket{\mb{a_{i}}}\ket{\mb{b_{j}}}\pm
    \ket{\mb{a_{j}}}\ket{\mb{b_{i}}} \;\mbox{,}
    \label{eq:pij2}
\end{eqnarray}
where we have introduced the normalized states
$\ket{\psi^{ij}}\propto(\ket{i}\ket{j}\pm\ket{j}\ket{i})$, and
$\mb{a_{i}}$ and $\mb{b_{i}}$ are the $i$th columns of 
$A^*$ and $B^*$ respectively. For b-qudits, double-clicks,
i.e. $i=j$, correspond to separable POVM elements,
while for f-qudits, the Pauli exclusion principle prohibits these 
events. In the last equations we also see that each $\ket{P^{ij}}$
has at most Schmidt rank\cite{terhal00} two. All other accessible 
POVM elements are convex combinations of those defined in 
Eq.~\eqref{eq:rk}. From here it follows that 
all POVM elements on two i-qudits realized with 
linear elements will have a Schmidt number\cite{terhal00} at most equal 
to two \cite{Aux}. 
This means that no analog of the incomplete Bell-measurement 
\cite{weinfurter94} \textemdash or its usage in 
teleportation\cite{bouwmeester97a}, entanglement swapping\cite{zukowski93},
quantum dense coding \cite{mattle96} or probabilistic implementation of
non-local gates \cite{dur01}\textemdash
can exist for i-qudits with $d>2$. Also from this result and 
a theorem by Carollo and Palma \cite{carollo} it follows\cite{dusek01} that 
even with the aid of auxiliary photons and conditional dynamics,
it is not possible to do a never failing Bell-measurement 
for qudits.

For $d=2$ it has been recently proven\cite{calsamig01}
that an incomplete Bell-measurement can at most unambiguously discriminate 
a Bell-state in half of the trials.
The appeal of a Bell-measurement is not only its ability
to discriminate unambiguously between the specific four Bell-states,
but that it can project an unknown state into a maximally 
entangled state. Any generalized measurement in which every 
POVM element is maximally entangled, would have much the same appeal.
Trivial modifications, consisting only in local operations,
could make the teleportation, entanglement swapping
 or the probabilistic non-local
 gates  function with such
 generalized Bell-measurement.
It is characteristic of bipartite maximally entangled pure states,
that each of its subsystems has a reduced density matrix 
proportional to the identity matrix. By
Eq.~\eqref{eq:red}, this means that, in the matrix representation,
maximally entangled states are proportional to unitary matrices. 
For both, bosons and fermions, 
the POVM elements defined in Eq.~\eqref{eq:pij2} that have a contribution
from the detection in a mode $c_{i}$ can be written, using 
a local base transformation, as 
\begin{equation}
   \ket{\tilde{P}^{ij}}\equiv W_{i}^\dagger\!\otimes\!V_{i}^\dagger \ket{P^{ij}}  =  
    |\mb{a_{i}}|\ket{1}\ket{\mb{x}}+
    |\mb{b_{i}}|\ket{\mb{y}}\ket{1} \mbox{,}
    \label{eq:pijtr} 
\end{equation}
where $\mb{x}= V_{i}\ket{\mb{b_{j}}}$ and $\mb{y}= W_{i} 
\ket{\mb{a_{j}}}$, and $V_{i}$ and $W_{i}$ are unitary transformations.
The matrix representation of this state is,
\begin{equation}
    \tilde{P}^{ij}=\left(
    \begin{array}{cc}
        |\mb{a_{i}}|x_{1}+|\mb{b_{i}}|y_{1} & |\mb{a_{i}}|x_{2}  \\
        |\mb{b_{i}}|y_{2} & 0
    \end{array}
    \right)\mbox{.}
    \label{eq:mpij}
\end{equation}
The following conditions have to hold if this matrix ought to be 
unitary (up to a constant $\kappa_{i}$),
\begin{equation}
    |\mb{a_{i}}|x_{1}+|\mb{b_{i}}|y_{1}=0 \;\mbox{ and }\;
    |\mb{a_{i}}||x_{2}|=|\mb{b_{i}}||y_{2}|=|\kappa_{i}|\mbox{.}
    \label{eq:conun}
\end{equation}
Enforcing these conditions, we have 
$\tilde{P}^{ij} =  \kappa_{i} \left({ 0 \atop {\mathrm e}^{i\phi}}
     {1 \atop 0}\right)$, which, after switching back to the state 
representation, allows us to conclude that
a detection in mode $c_{i}$ can only contribute
to maximally entangled POVM elements of the form
\begin{equation}
         \ket{P^{ij}} =  \kappa_{i} W_{i}\!\otimes\!V_{i}
     (\ket{1}\ket{2}+\mbox{e}^{i \phi}\ket{2}\ket{1})\mbox{.}
    \label{eq:maxentg}
\end{equation}
On the other hand, after some simple algebra and using $AB^\dagger=0$,
one can find the total contribution, in the resolution of the identity,
of all the POVM elements where a detection in the $c_{i}$ mode is involved, 
\begin{eqnarray}
 \id=\sum_{i\geq j=1}^{n}\ketbrad{P^{ij}}  &=&
  \sum_{i=1}^{n} \frac{\kappa_{i}}{2}
    W_{i}\!\otimes\! V_{i}\left(|\mb{a_{i}}\!|^2\ketbrad{1}\!\otimes\!
    \id_2\right.
    \nonumber \\
   &+&\left.|\mb{b_{i}}\!|^2 \id_{2}\!\otimes\!\ketbrad{1}\right)
   W_{i}^\dagger\!\otimes\! V_{i}^\dagger\label{eq:sum2}
\end{eqnarray}
The factor $\frac{1}{2}$ comes from
the symmetry $P^{ij}\!=\!P^{ji}$ and compensates the double counting
of the terms with $i\!\neq\!j$.
Comparing this result with Eq.~\eqref{eq:maxentg} it is clear
that not all POVMs involving a detection in $c_i$ can
be maximally entangled; the space spanned by POVMs
defined by Eq.~\eqref{eq:maxentg} does not cover
the whole support of the \emph{i}th term in the sum in
Eq.~\eqref{eq:sum2}.
An upper bound on the total weight of the maximally entangled POVMs
in this term fixes the maximum probability of successfully projecting
an unknown input state $\rho=\frac{1}{4}\id_{2}\!\otimes\!\id_{2}$ onto 
a maximally entangled state,
\begin{eqnarray}
 &&p^i_{{\mathrm succ}}\leq \frac{1}{2}\Tr\left( W_{i}\!\otimes\! V_{i}
    (|\mb{a_{i}}\!|^2\ketbrad{1}\!\otimes\!\ketbrad{2}
    +|\mb{b_{i}}\!|^2\ketbrad{2}\!\otimes\!\ketbrad{1}) \right. \nonumber \\
     &&
   \left. W_{i}^\dagger\!\otimes\! V_{i}^\dagger\rho\right)
    =\frac{1}{8}(|\mb{a_{i}}\!|^2+|\mb{b_{i}}\!|^2)=\frac{1}{8} 
    \sum_{k=1}^{4} |U_{ik}|^2\mbox{,}
    \label{eq:psucc}
\end{eqnarray}
where we employed the definition in Eq.~\eqref{eq:Uabc}.
By adding up the contributions from all the detectors we obtain
total probability of success,
\begin{equation}
    \sum_{i=1}^{n} p^i_{{\mathrm succ}}\geq \frac{1}{8} 
    \sum_{k=1}^{4}\sum_{i=1}^{n} |U_{ik}|^2= \frac{1}{8} 
    \sum_{k=1}^{4}1=\frac{1}{2}\mbox{.}
    \label{eq:totalfail}
\end{equation}
This also sets to one half the ultimate efficiency of teleportation,
entanglement swapping and the 
probabilistic implementation of non-local gates on i-qubits.

In this communication I have introduced a formalism to
study the characterization of the generalized measurements
on two i-qudits implemented by linear elements.
This approach provides two otherwise
non-trivial results concerning maximally entangled
POVMs. It should also be very helpful in 
determining the viability or efficiency of 
other relevant POVM in quantum information.

As a last remark, even if the results for fermions and bosons
are apparently similar, there is actually a large difference
that is manifest in the asymptotic method that also uses
auxiliary photons. In Ref.\cite{knill01b} it is proved
that the analogue of the photonic efficient quantum computation 
\cite{knill01} can not exist for fermions.

I would like to thank N. L\"utkenhaus, K.-A. Suominen, and M. Mackie
for reading the manuscript and M. Du{\v s}ek for the motivation
on Bell measurements for qudits\cite{dusek01}.
This work was supported by the Academy of Finland (project 43336)
and the EU IST EQUIP program.



\begin{thebibliography}{10}
\bibitem{Aux} 
It is a straightforward extension of this result 
that any ``click'' pattern 
corresponds to a POVM element with Schmidt rank
smaller or equal to the number of involved particles
(i-qudits$+$auxiliary).
\end{thebibliography}

\end{multicols}

\end{document}